\documentclass[journal]{IEEEtran}
\usepackage{booktabs}
\usepackage{subfigure}
\usepackage{cite}
\usepackage{amsmath,epsfig,bm}
\usepackage{pifont}
\usepackage{graphicx}
\usepackage{amssymb}
\usepackage{amsmath}
\usepackage{cancel}
\usepackage[normalem]{ulem}
\usepackage[dvips]{color}
\usepackage{caption}
\usepackage{lineno}
\usepackage{citesort}

\newcommand{\Hnull}{\mathcal{H}_0}
\newcommand{\Halt}{\mathcal{H}_1}
\newcommand{\Honull}{\mathcal{{D}}_0}
\newcommand{\Hoalt}{\mathcal{{D}}_1}
\newcommand{\Hini}{\mathcal{H}_i}
\newcommand{\Houtj}{\mathcal{D}_j}

\title{Location Verification Systems in Emerging \\Wireless Networks}
\author{
\IEEEauthorblockN{Shihao Yan,
                  Robert Malaney\\
\IEEEauthorblockA{
School of Electrical Engineering and Telecommunications, UNSW, Sydney, Australia}}
}

\begin{document}


\maketitle

\begin{abstract}
As location-based techniques and applications become ubiquitous in emerging wireless networks,  the verification of location information will become of growing importance. This has led in recent years to an explosion of activity related to location verification techniques in wireless networks, with a specific focus on Intelligent Transport Systems (ITS) being evident. Such focus is largely due to the mission-critical nature of vehicle location verification  within the ITS scenario. In this work we review recent research in wireless location verification related to  the vehicular network scenario. We particularly focus on location verification systems
that rely on formal mathematical classification frameworks, showing how many systems are either  partially or fully encompassed by such frameworks.
\end{abstract}

\section{Introduction}\label{introduction}

 As location-based techniques and services become ubiquitous in emerging wireless networks, the authentication of location information  has attracted considerable research interest in recent years \cite{malaney2004location,raya2005the,faria2006detection,papad2006securing,raya2007securing,sheng2008detecting,zhang2008evaluation,papad2008secure,bao2008location,bauer2009the,zekavat2011handbook,yang2013detection}. In early wireless positioning systems, accuracy and performance issues were to the fore, with the authentication of location information relegated to a secondary concern. This is now changing. Many current mainstream wireless positioning systems, such as the now ubiquitous WiFi positioning systems,  are highly vulnerable to location-spoofing attacks due to their openness and wide public availability \cite{ledvina2004bitwise,tippen2008iphone}. In particular, in many configurations wireless network positioning systems are  client-based\cite{hofmann1997global,bulusu2000gps}, meaning that only the client (the device whose location is to be verified) can obtain its location directly. The wider communications network can then only obtain the client's position through requesting the client to report its location. Obviously, the client can easily spoof  or falsify its  location. In other configurations, systems that attempt to directly locate a client using signal metrics, such as  received signal strength (RSS) measurements, are vulnerable to manipulation of the signal metric by the client prior to transmission \cite{li2005robust,capkun2006secure,liu2008attack,mahfouz2013distance}.

 It is the purpose of this paper to review those works that attempt to formalize location spoofing attempts. We will be focussed on the notion of \emph{location verification}, meaning that we will assume a location (presumed to be the actual true location) for the client is either publicly announced by the client or is assumed to be \emph{a priori} publicly known. We refer to this announced (or known) location as the \emph{claimed location}. The verification systems  we discuss are then instructed to use all available signal metrics in order to classify whether the client is at the claimed location or not [21-54]. It is important to note that location verification (or authentication) defined in this way results a quite different mathematical problem (and different outcome) to that posed by the more usual location acquisition problem \cite{hofmann1997global,bulusu2000gps,li2005robust,capkun2006secure,liu2008attack,mahfouz2013distance}.

The importance of location verification can be witnessed by the many adverse effects spoofed location information can have on a variety of network functions \cite{raya2005the,papad2006securing,raya2007securing,papad2008secure,leinmuller2005influence,leinmuller2006greedy,mauve2001survey,yang2008connectivity,rabayah2012new,bertino2005geo,chen2006inverting,capkun2010integrity}.
 For example, in generic wireless  network scenarios, spoofed position information can lead to packet delivery in geographic  routing protocols \cite{mauve2001survey,yang2008connectivity,rabayah2012new} being reduced dramatically  \cite{leinmuller2005influence,leinmuller2006greedy}.  Performance of location-based access control can be decreased markedly by spoofed locations \cite{bertino2005geo,chen2006inverting,capkun2010integrity}. As mentioned in \cite{malaney2006secure}, WiFi, Cellular and  GPS position information within the E911 framework can be easily spoofed by clients,
 in order to maliciously attract emergency services to false locations.
However, the adverse effects of location spoofing are arguably more severe in the vehicular ad hoc network (VANET) scenario \cite{raya2005the,papad2006securing,raya2007securing,papad2008secure}. Most importantly, in the collision avoidance aspects of VANETs location spoofing can be life-threatening. Beyond such critical effects,  a malicious  vehicle might spoof its position  in order to cause serious service disruptions to other users \cite{raya2005the,papad2006securing,raya2007securing}, or to enhance in a selfish manner its own functionality within the network \cite{jaeger2012novel,yu2013detecting}. Authentication of position information within VANETs  forms the focus of the rest of this paper.

In this review paper, the approach which will gather most of our attention is the exploitation of the physical properties of wireless communication channels as a means to derive location verification. Such an approach eliminates (or at least drastically reduces) any dependency on complex higher-layer secrecy techniques such as encryption and cryptographic key management. Use of the   properties of the wireless communication channels also allows us to more formally examine what  the optimal performance expectations are for a Location Verification System (LVS).

The rest of this paper is organized as follows. Section
\ref{LVS} details a generic formal LVS and presents some associated performance evaluation criteria. In Section \ref{VANET} we review application of our generic  LVS  to the emerging VANET scenario.  In Section \ref{other} we discuss  other location verification systems that although not targeted directly at the VANET scenario - can be readily adjusted to work in that scenario. Section \ref{Conclusions} draws some concluding remarks and future directions.

\section{Generic Location Verification System}\label{LVS}

In this section, we first present the generic system model of an LVS, and discuss the differences between an LVS and a localization system. Then, the receiver operating curves (ROC) used in an LVS are discussed. Finally, we present two frameworks for optimizing an LVS.

\subsection{Binary Decision Rule for an LVS}\label{LVS_decision_rule}

\begin{figure}[!t]
\begin{center}
{\includegraphics[width=3in]{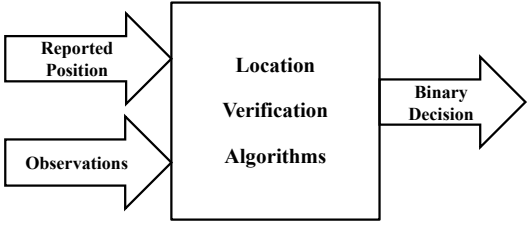}}
\end{center}
\caption{Illustration of a generic location verification system.}
\label{fig:LVS}
\end{figure}

Location verification is different from the more general problem of locating a user in wireless networks \cite{malaney2007wireless}. A key obvious difference between an LVS and a positioning system is that the output of an LVS is usually a binary decision (yes/no) while the output of a positioning system is an estimated location of a user. Fig. \ref{fig:LVS} depicts a generic location verification system. Here, the \emph{claimed location} is an LVS input that is provided by the client being verified, whom henceforth we refer to as the \emph{prover}. The observations available at the \emph{verifiers}, which perform location verifications, are also  inputs to the LVS.  The LVS aims at verifying a prover's claimed location by comparing these inputs and checking in a systematic manner whether all inputs are compatible. If compatible (incompatible), a yes (no) decision is returned by the LVS.

Due to the binary outputs, location verification can be modeled as a binary decision theory problem. Thus, a decision rule is embedded in an LVS, which can be written as
\begin{equation}\label{decision}
T \begin{array}{c}
\overset{\Hoalt}{\geq} \\
\underset{\Honull}{<}
\end{array}%
\lambda,
\end{equation}
where $T$ is the test statistic, $\lambda$ is the threshold corresponding to $T$, and $\Honull$ and $\Hoalt$ are the binary decisions inferring the prover of interest is legitimate or malicious, respectively. The test statistic $T$ is derived from the LVS inputs, the form of which varies with different location verification algorithms.

 As mentioned above, the principal aim of an LVS is to verify if a prover is at the claimed location. Of course we have to  include noise (for example, that in the location acquisition problem that results in location error) in such a framework, but that is usually encapsulated within the classifier logic (i.e. implicit probability distributions) of the LVS. In addition, if the verification was changed from a  claimed \emph{position} to a claimed \emph{area} (\emph{e.g.} a room) then the systems implicit probability distributions would be altered accordingly.

\subsection{ROC's for an LVS}\label{LVS_criteria}

\begin{figure}[!t]
\begin{center}
{\includegraphics[width=2.7in]{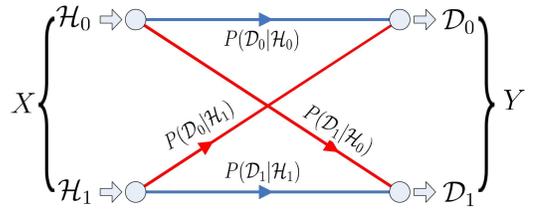}}
\end{center}
\caption{Statistical decision theory model for an LVS\cite{yan2012optimal}.}
\label{fig:LVS_high}
\end{figure}

From a statistical decision theory point of view, an LVS can be modeled as a decision problem as shown in Fig.~\ref{fig:LVS_high}. In this figure, $X$ denotes the input of an LVS, of which the two realizations are $\Hnull$ and $\Halt$. Here $\Hnull$ represents the case where the prover is at the claimed location (null hypothesis), and
and $\Halt$ represents a malicious prover not at the claimed location (alternative hypothesis). As shown in Fig.~\ref{fig:LVS_high}, the output of an LVS, $Y$, is also binary, and its realizations are $\Honull$ and $\Hoalt$. The probabilities  $P(\Houtj|\Hini)$ ($i = 0,1$, $j = 0, 1$) represent for a given input $X$ the probabilities of deciding on an output $Y$.  The  $P(\Houtj|\Hini)$  determine the performance of an LVS ($e.g.$ \cite{yan2012optimal}).

The traditional method to evaluate a detection system utilizes the Neyman-Pearson Lemma \cite{neyman1933problem,barkat2005signal}, which in turn is based on the false positive and detection rates. The false positive rate is the probability of deciding incorrectly that a legitimate prover is malicious, denoted by $\alpha = P(\Hoalt|\Hnull)$. The detection rate is the probability of deciding correctly that a prover is malicious, which is presented as $\beta = P(\Hoalt|\Halt)$. Both the aforementioned rates are determined by $T$ and $\lambda$.
We are expecting an LVS to provide a high detection rate and a low false positive rate. But there is a tradeoff between the false positive and detection rates. The ROC is used to demonstrate this tradeoff, and is constructed by plotting $\beta$ versus $\alpha$. But ROC in and by itself does not lead to an optimized setting ($e.g.$ \cite{yan2012optimal}).

It is worth noting that the Neyman-Pearson Lemma states that the \emph{likelihood ratio test} is able to maximize the detection rate for any given false positive rate \cite{neyman1933problem,barkat2005signal}. The likelihood ratio is the ratio between the probability density functions of the measurements under $\Hnull$ and $\Halt$, respectively, and the corresponding threshold $T$ is derived assuming a given false positive rate. However, determining the threshold $T$ by such a method does not provide for an overall optimization of an LVS ($e.g.$ \cite{neyman1933problem}, \cite{barkat2005signal}).


\subsection{Optimization Frameworks for an LVS}\label{LVS_optimization}

To optimize an LVS, some unique evaluation criterion should be adopted as the performance metric. As mentioned above the transition probabilities between $X$ and $Y$ determine the performance of an LVS, thus the unique evaluation criterion should be a function of such transition probabilities.
One widely used metric is the Bayes's average cost, which is defined as\cite{barkat2005signal},
\begin{equation}\label{Bayes}
\mathfrak{R}= \sum_{i=0}^{1} \sum_{j=0}^{1} C_{ji} P(\mathcal{D}_j|\mathcal{H}_i) P(\mathcal{H}_i),
\end{equation}
where $C_{ji}$ is the assigned cost associated with the decision $\mathcal{D}_j$ given the hypothesis $\mathcal{H}_i$, and $P(\mathcal{H}_0)$ and $P(\mathcal{H}_1)$ are the \emph{a priori} probabilities of the occurrence of $\Hnull$ and $\Halt$, respectively. In this Bayesian framework, the optimal location verification algorithm is the one which minimizes $\mathfrak{R}$.  It is worthwhile to note the Bayes's average cost requires that both $C_{ji}$ and $P(\mathcal{H}_i)$ are known \emph{a priori}.

If the $C_{ji}$'s are unknown, the MAP criterion can be adopted \cite{barkat2005signal}, where the unique cost is determined through
\begin{equation}\label{MAP}
\mathfrak{R}_M= C_{10} P(\mathcal{D}_1|\mathcal{H}_0)P(\mathcal{H}_0) + C_{01} P(\mathcal{D}_0|\mathcal{H}_1)P(\mathcal{H}_1).
\end{equation}
Comparing \eqref{Bayes} and \eqref{MAP}, we can see that $\mathfrak{R}_M$ is a special case of $\mathfrak{R}$ with $C_{00} = C_{11} = 0$, and $C_{10} = C_{01} = 1$. Thus, the MAP criterion is well suited for scenarios where the cost of rejecting a legitimate user is equal to that of accepting a malicious user. In addition, $\mathfrak{R}_M$ can be reduced to the \emph{Total Error} when $P(\mathcal{H}_0) = P(\mathcal{H}_1) = 0.5$ \cite{chen2010detecting}.


Although the Bayes's average cost, MAP,  and Total Error can be used to optimize an LVS, they all possess a weakness - they are dependent on \emph{subjective} cost values. In the Bayes's average cost we have to predetermine the costs for all possible decisions. However, properly determining the true cost to the network for each decision is in practice next to impossible. For example, what is the detailed cost to a VANET where one vehicle spoofs its location information? Clearly that true cost is subjective and dependent on numerous factors and application scenarios.
Similarly, $C_{10}$ and $C_{01}$ are both set to one in the MAP criterion and Total Error, but the cost of accepting a malicious prover is likely much larger than that of rejecting a legitimate prover.

To solve this subjective problem in the optimization of an LVS, an information-theoretic framework has been proposed in \cite{yan2012information,yan2012optimal}, where the cost for each possible decision is not assigned subjectively, but rather \emph{objectively}.
In this framework, the mutual information between the system input and output is utilized as the optimization criterion. The mutual information between $X$ and $Y$ is defined as
\begin{equation}
I(X;Y)= H(X) - H(X|Y),
\end{equation}
where $H(X)$ is the input entropy representing the uncertainty of the system input (determined by the \emph{a priori} probabilities), $H(X|Y)$ is the conditional entropy of $X$ given $Y$, which measures the uncertainty of the input given the output (determined by the \emph{a priori} and transition probabilities). Thus, $I (X;Y)$ measures the uncertainty reduction of the input given the output, and the optimal information-theoretic location verification algorithm is the one which maximizes $I (X;Y)$. Comparing with the Bayesian framework, the information-theoretic framework only assumes knowledge of the \emph{a priori} probabilities.


In closing this section we note the following important point. Although the likelihood ratio has been shown to be the optimal test statistic in most frameworks, this ratio is difficult to obtain in practice without some assumptions on an attacker's behavior and location. The reason for this is that the likelihood function under $\Halt$ is directly dependent on the attack strategy adopted by the malicious prover and the actual (true) location of the malicious prover, both of which are unknown \emph{a priori} to the LVS. The impact of these key uncertainties with regard to optimization of the LVS forms part of the following discussions.

\section{Location Verification in VANETs}\label{VANET}

In this section we review  applications of our generic LVS framework to the emerging VANETs scenario. Firstly we review works that only discuss the binary decision rule of Section~\ref{LVS_decision_rule}. Secondly, we review works that also use the ROC's of Section~\ref{LVS_criteria}. Thirdly we review works that utilize the optimization frameworks discussed in Section~\ref{LVS_optimization}.

%


%
%

\subsection{Binary-Decision-Rule-Based Location Verification}


In this subsection, we focus on the location verification algorithms that only utilize the binary decision rule as discussed in Section~\ref{LVS_decision_rule}. In these algorithms the threshold is physically set to some system parameter ($e.g.$ maximum communication range, maximum velocity).

By exploiting the specific properties of VANETs, such as high node density and mobility, the authors of \cite{leinmuller2006position} proposed an autonomous scheme  and a cooperative scheme to detect and mitigate falsified locations. The acceptance range, mobility grade and vehicle density are used in the binary decision rule in the autonomous scheme, where the thresholds are determined based on the maximum communication range, maximum velocity and maximum density, respectively. The test statistics used in the cooperative scheme, such as neighbor tables, can only be obtained through cooperating with other neighbor vehicles. The overall decision on a prover's claimed location is made by combining the local decisions with weight factors. Since the proposed location verification is applied in location-based routing protocols, in which it is assumed a malicious vehicle does not forward the packet to the correct next hop, the packet delivery ratio can be used as an performance criterion. The proposed schemes in \cite{leinmuller2006position} provide the basis of location verification in VANETs. Similar to \cite{leinmuller2006position}, the authors of \cite{alsharif2011mitigating} also proposed location verification algorithms based on communication range, velocity and density, but extended their test statistics to include  travelled distance and map location.

The authors of \cite{harsch2007secure} used the timestamp of when a packet was sent (within which the claimed location is embedded) to detect malicious vehicles (\emph{i.e.} vehicles who spoof location information) in VANETs. The timestamp check is to ensure the received packet is neither too old nor too early. In \cite{harsch2007secure}  a rate-limiting mechanism was also proposed. In this mechanism if the rate of packets originating from a prover exceeds a predetermined maximum packet transmit rate, the prover will be detected as malicious.  As in \cite{leinmuller2006position}, the work of \cite{harsch2007secure} is in the context of location-based routing, and  the packet delivery ratio could have been used as a performance criterion. However, the performance metric used in  \cite{leinmuller2006position} was packet end-to-end delay, which is also a viable metric when it is assumed a malicious vehicle does not forward the packet to the correct next hop.

A secure, infrastructure-less, and cooperative location verification scheme was proposed in \cite{song2008secure} with the aim of preventing an attacker from spoofing that it is further from a verifier. In this scheme, the verifier first estimates the prover's location based on time difference of arrival (TDOA) measurements, with the help of a common neighbor (of both the verifier and prover). Then, the Euclidean  distance error between the estimated and claimed locations of the prover is compared with a distance related to the expected processing delay (assuming a legitimate prover). The packet delivery ratio and  end-to-end delay are applied to evaluate the proposed location verification algorithm, since (again) the work of \cite{song2008secure} is in context of location-based routing protocols.

\begin{table}
\caption{Binary decision rule based location verification algorithms for VANETs}
\begin{center}
\begin{tabular}{c | c | c }
\hline
Ref. & Test Statistics & Performance Criteria \\
\hline
\cite{leinmuller2006position} & \parbox[c]{3.5cm}{acceptance range, mobility grade, maximum density, neighbor tables, $etc.$} & \parbox[c]{3.5cm}{packet delivery ratio} \\
\hline
\cite{harsch2007secure} & \parbox[c]{3.5cm}{timestamp, acceptance range, velocity, packet transmit rate} & \parbox[c]{3.5cm}{packet end-to-end delay}\\
\hline
\cite{yan2008providing} & \parbox[c]{3.5cm}{error distance between radar estimated location and claimed location} & \parbox[c]{3.5cm}{time required to detect a malicious vehicle}\\
\hline
\cite{song2008secure} & \parbox[c]{3.5cm}{Euclidean distance error between estimated and claimed locations} & \parbox[c]{3.5cm}{packet delivery ratio, packet end-to-end delay}\\
\hline
\cite{yan2009providing} & \parbox[c]{3.5cm}{difference between track records and claimed positions, \\neighbor tables} & not provided\\
\hline
\cite{xue2010trusted} & \parbox[c]{3.5cm}{TOA of challenge-response message, acceptance range, roadway map, velocity} & area of intersection region\\
\hline
\cite{alsharif2011mitigating} & \parbox[c]{3.5cm}{communication range, speed and density, moved distance} & {packet delivery ratio}\\
\hline
\cite{weer2011verifying} & \parbox[c]{3.5cm}{difference between measured and calculated round trip time} & \parbox[c]{3.5cm}{difference between measured and calculated round trip time and distances}\\
\hline
\cite{abumansoor2012secure} & \parbox[c]{3.5cm}{difference between triangulation calculated and claimed distances} & \parbox[c]{3.5cm}{channel capacity utilization, packet delivery ratio, response time}\\
\hline
\cite{das2013position} & \parbox[c]{3.5cm}{Euclidean distance error between estimated location based on TDOA measurements and claimed location} & \parbox[c]{3.5cm}{average location estimation error}\\
\hline
\end{tabular}
\end{center}
\label{table_decision}
\end{table}

In \cite{yan2008providing} the authors utilized on-board radar systems to verify a vehicle's claimed location  (obtained through a GPS system). Considering the system noise, the authors first determined the GPS position tolerance shadow and radar position tolerance shadow, separately.  The proposed algorithm accepts the prover's claimed location if there is an intersection between the GPS and radar position shadows, and vice versa. The threshold and performance of the proposed location verification algorithm is determined by the accuracy of the GPS and radar systems. The time required to detect a malicious user is used as the evaluation criterion. Again, the work of \cite{yan2008providing} is in the context of location-based routing protocols, and packet end-to-end delay and delivery ratio are also used. The authors of \cite{yan2008providing} also proposed in \cite{yan2009providing} a passive location verification algorithm which can work in scenarios where the on-board radar is not available (or does not work due to obstacles). This passive algorithm first creates a track record of location reports by using neighbor tables. A vehicle's neighbor table contains a list of other vehicles' identifications and locations that are within its range. If a prover's claimed location greatly deviates from the majority of the track record, the algorithm will classify the prover as malicious.

A location verification algorithm for VANETs in the context of location-based routing was proposed in \cite{xue2010trusted} based on the notion of \emph{trusted} neighbor (a vehicle whose location has been verified). The proposed algorithm is carried out in two steps. In the first step, the time of arrival (TOA) of the challenge-response message between a verifier and a prover is utilized to detect distance reduction attacks. In the second step, the verifier cooperates with one of its trusted neighbors to verify if the prover is at an intersection region determined by the verifier and the trusted neighbor. The area of the intersection region is used as the performance criterion.

In \cite{weer2011verifying} a scheme, in which a verifier  measures the round trip propagation time between the verifier and the prover (by exchanging challenge-response messages), and  compares this  with a calculated round trip time (obtained according to the prover's claimed location) is studied via detailed numerical simulations. This work is notable in which it concludes that such easily deplorable schemes are seemingly reliable in rural, urban and Manhattan scenarios.

To overcome the no line-of-sight (LOS) problem in location verification systems, a cooperative location verification scheme was proposed in \cite{abumansoor2012secure}. The proposed scheme focused on verifying a prover with no LOS to a verifier. To estimate the distance between the prover and verifier, this protocol requests help from a cooperative vehicle, which has LOS communications with both the prover and the verifier. The distances from the cooperator to the prover, and from  the cooperator to the verifier, can be estimated  ($e.g.$ via TDOA, or  TOA), which then allows for the distance between the prover and verifier to be calculated. In addition, the claimed distance between the verifier and the prover's claimed location can be calculated. The main point of this protocol is its ability to verify vehicle locations that could not otherwise be verified due to obstacles.

The authors of \cite{das2013position} proposed a location verification algorithm dedicated to VANETs, where a moving verifier (vehicle) can verify a static prover's claimed location independently of road-side units and other neighbor vehicles. In this algorithm, the moving verifier  measures the TOA of signals transmitted by the static prover at three different locations along its trajectory. Then, the location of the prover is estimated based on the three measurements using the multilateration technique. The Euclidean  distance error between the estimated and claimed locations of the prover is used as the test statistic, and the corresponding threshold is set as the average position estimation error.

The detailed test statistics and performance criteria used in the location verification algorithms reviewed in this subsection are provided in Table~\ref{table_decision}.

\subsection{Location Verification Using ROC's}

In this subsection, we will review the location verification systems, not only utilizing the binary decision rule, but also using  both or either of the false positive rate and detection rate (used to construct a ROC) discussed in Section~\ref{LVS_criteria}. In these location verification systems, the threshold in the binary decision rule could be set according to a given false positive rate or detection rate.

%
%

A location verification solely based on messages exchange among neighbor vehicles was proposed in \cite{ren2009location}. The authors focused on detecting a malicious vehicle which spoofs its position as the farthest one (within range) from the packet sender, so that it will be selected as the next hop in geographic routing protocols. In  \cite{ren2009location} it is assumed each vehicle is equipped with two directional antennas: forwards and backwards, and each vehicle constructs two corresponding  tables of one-hop neighbors. The decision on a prover is made though exchanging and comparing such neighbor tables. The theoretic detection rate is derived as a function of the vehicle density. As expected, it is found that the larger the network density, the higher that malicious vehicles will be detected by the proposed system.

 A location verification algorithm based on a vehicle's direct connectivity (one-hop connectivity) with other vehicles was proposed in \cite{abu2011position}. In this algorithm  one-hop information is exchanged between vehicles  so that  each vehicle can build  a two-hop neighborhood connectivity diagram. Using such diagrams, each vehicle can then attempt to verify the location information being passed to it. Each vehicle does this by constructing a \emph{plausibility area}. Simply put, if a vehicle cannot hear directly from another vehicle, say vehicle A, at some location (since that vehicle is two hops away), then it should not be able to hear directly from a prover who claims to be further away than vehicle A. Similarly, in \cite{abu2012map} a map-guided trajectory-based location verification algorithm was proposed in which the  plausibility area is constructed by using a prover's history location and map information (\emph{e.g.} road dimensions).

\begin{table}
\caption{Location verification algorithms using Neyman-Pearson criteria for VANETs}
\begin{center}
\begin{tabular}{c | c | c }
\hline
Ref. & Test Statistics & Performance Criteria \\
\hline
\cite{ren2009location} & \parbox[c]{3.5cm}{number of neighbor vehicles} & \parbox[c]{3.5cm}{ detection rate, packet delivery ratio, packet end-to-end delay}\\
\hline
\cite{abu2011position} & \parbox[c]{3.5cm}{two-hop neighbors based plausibility area, RSS measurements} & \parbox[c]{3.5cm}{ detection rate, false positive rate}\\
\hline
\cite{abu2012map} & \parbox[c]{3.5cm}{map-guided trajectory based plausibility area, RSS measurements} & \parbox[c]{3.5cm}{ detection rate, false positive rate}\\
\hline
\cite{zhang2012cooperative} & \parbox[c]{3.5cm}{Difference between estimated distance based on TDOA and calculated distance based on claimed location} & \parbox[c]{3.5cm}{detection rate, packet delivery ratio}\\
\hline
\cite{chen2013beacon} & \parbox[c]{3.5cm}{overall trustworthiness of a message} & \parbox[c]{3.5cm}{false positive rate, detection rate, false negative rate, true negative rate}\\
\hline
\cite{yu2013detecting} & \parbox[c]{3.5cm}{error distance between estimated and claimed locations, variance of this error distance} & \parbox[c]{3.5cm}{false positive rate,  detection rate}\\
\hline
\end{tabular}
\end{center}
\label{table_criteria}
\end{table}

In order to prevent the distance enlargement attack in VANETs, the authors of  \cite{zhang2012cooperative} also proposed a cooperative verification algorithm to verify a prover's claimed location. In this scheme, both the verifier and cooperator can measure the TOA of the challenge-response messages from a prover. By using such TOA measurements, both the verifier and cooperator can conduct local verification on whether the prover launched distance reduction attacks. In such location verification algorithms, the test statistic is the difference between the TOA calculated distance and claimed distance derived from the prover's claimed location, and the threshold is determined using the processing delay of the challenge-response message. Since the cooperator is selected so as to locate the prover between the verifier and cooperator, the proposed cooperative algorithm is able to detect the distance enlargement attack.  In the simulations of \cite{zhang2012cooperative}, the detection rate is used to evaluate the proposed LVS.

 In contrast to the previously reviewed works which focus on the one-hop location verification problem, the authors of \cite{chen2013beacon} proposed a beacon-based trust management system, which combines the one-hop and multiple-hop verification algorithms to thwart internal attackers in VANETs. In the proposed system, the authors adopted the cosine similarity \cite{yan2008providing} between estimated vector (including position and velocity) and claimed vector in order to determine the beacon trustworthiness of a neighbor vehicle. The Tanimoto  coefficient between history beacon messages and received event messages is utilized to calculate the one-hop event trustworthiness, based on which an algorithm to determine the multiple-hop trustworthiness of an event message is also provided. Then, the Dempster-Shafer theory \cite{chen2005dempster} is applied to combine all local event trustworthiness and determine the overall trustworthiness of an event message. Finally, the overall decision on the beacon message is made by comparing the overall trustworthiness with a predetermined threshold of trust degree. Both the false positive and detection rates are utilized as performance criteria.

Besides location spoofing attacks, Sybil attacks may also compromise some location-based services in VANETs. The Sybil attack refers to the scenario where a malicious vehicle illegitimately adopts multiple identities or locations to realize its attack purposes. This type attack is possibly launched by a selfish driver to mimic traffic congestion at some location on the road (used say as a mechanism to deter other vehicles from driving into his planned path). To detect such Sybil attacks, two location verification algorithms based on RSS measurements were proposed in \cite{yu2013detecting}. In the first algorithm of \cite{yu2013detecting}, the verifier first estimates the prover's location through the Minimum Mean-Square Error on the distribution of RSS measurements. Then, the distance error between the estimated and claimed locations is utilized as the test statistic. In the second  algorithm, the test statistic is derived from the distributions of such distance errors under $\Hnull$ and $\Halt$, and the threshold is derived from a given false positive rate. In the conducted simulations, \cite{yu2013detecting} adopted the detection rate as the performance criterion.




\subsection{Information-Theoretic Location Verification Systems}

In this subsection we will discuss some results from the information-theoretic optimization framework  \cite{yan2012information,yan2012optimal} discussed in Section~\ref{LVS_optimization}. We then discuss the advantages of the information-theoretic framework relative to the Bayesian framework, in the context of location verification.


In \cite{yan2012information}, the verifier first uses the Maximum Likelihood Estimator to estimate a prover's location based on RSS measurements. The Mahalanobis distance error between the estimated and claimed positions of the prover is utilized as the test statistic in the binary decision rule. The corresponding threshold is selected by maximizing the
mutual information between the input and output of the LVS. It is worth noting that a threshold selected in this manner does not completely optimize an LVS since the test statistic is not optimized.

Following \cite{yan2012information}, the authors of \cite{yan2012optimal} proved that the likelihood ratio is the optimal test statistic in terms of maximizing the mutual information between input and output of an LVS. With such a test statistic an optimal information-theoretic LVS was obtained. To deploy the optimal information-theoretic location verification, \cite{yan2012optimal} also proposed three threat models, where the likelihood functions for RSS measurements can be obtained or approximated in closed forms. It is worth noting that the information-theoretic and MAP frameworks lead to the same decision rule (likelihood ratio test with the same threshold) in the special case where $P(\mathcal{H}_0) = P(\mathcal{H}_1) = 0.5$, and the likelihood functions under $\Hnull$ and $\Halt$ follow Gaussian distributions with the same variance.

Besides the objective nature of the optimization metric discussed in Section~\ref{LVS_optimization}, the authors of \cite{yan2012optimal} also showed another useful property of information-theoretic framework - the optimized threshold is not very sensitive to the \emph{a priori} probabilities, as shown by Fig.~9 in \cite{yan2012optimal}. From this figure, we can see that the optimal threshold in the MAP framework which minimizes $\mathfrak{R}_M$ for the likelihood ratio test is a linear function of $P(\Halt)$. This can be explained by the fact that the optimal threshold for minimizing $\mathfrak{R}_M$ is $P(\Hnull)/P(\Halt)$ \cite{barkat2005signal}. However, the optimal threshold in the information-theoretic framework that maximizes $I(X;Y)$ for the likelihood ratio test converges to a constant number as $P(\Halt)$ approaches zero. It is important to note that  knowledge on the \emph{a priori} probabilities is very difficult to obtain, and in practice can only be assumed or roughly estimated. In most practical verification scenarios one could anticipate $P(\Halt)$ being quite small.
Therefore, the behavior shown by Fig.~9 in \cite{yan2012optimal}, for the optimal information-theoretic threshold, provides information-theoretic frameworks a significant pragmatic advantage relative to Bayesian frameworks.



\section{Location Verifications Applicable to VANETs} \label{other}

In this section, we review some other location verification algorithms that, although not dedicated directly to VANETs,  can be readily adjusted to the VANET scenario.

Since collecting RSSs does not require extra hardware, many location verification systems for general wireless networks were developed based on RSS measurements. In \cite{faria2006detection}, the authors proposed an algorithm to detect location spoofing attacks by matching the input instantaneous measurements with the normal signal fingerprints. Through exploiting experimental test results, the authors of \cite{sheng2008detecting} found that the RSSs follow a mixture of two Gaussian distributions if the prover and verifier are both equipped with two antennas. To perform the verification, \cite{sheng2008detecting} employed a likelihood ratio test constructed from the instantaneous measurements and expected normal profiles. An RSS fingerprints based location verification algorithm was also proposed in \cite{malaney2007securing}, where it was observed that solely analyzing the residual of RSS measurements can not robustly detect location spoofing attacks. However, if this residual is used referenced to a claimed location, it can provide for a verification algorithm robust against various forms of attacks. In \cite{chen2010detecting}, the location verification was formulated as a statistical significance testing problem. The authors analyzed the spatial correlation of RSS measurements to detect location attacks, and derived theoretic false positive and detection rates in the 1-Dimension and 2-Dimension physical spaces. The authors of \cite{chen2010detecting} also optimized the threshold in the proposed binary decision rule by minimizing the Total Error, and proposed a location verification algorithm against spoofing and Sybil attacks  by using clustering methods algorithms \cite{hastie2001the}. The algorithms of \cite{faria2006detection} \cite{sheng2008detecting} \cite{chen2010detecting} \cite{malaney2007securing} are representative of many similar RSS based wireless local verification algorithms. In principal they could all be readily adopted to the VANET environment.

Some generic challenge-response based location verification algorithms for wireless networks have been proposed in the literature \emph{e.g.} \cite{sastry2003secure,capkun2008secure,zhang2008evaluation}.
The well-known Echo protocol was proposed in \cite{sastry2003secure} and is based on the delay of the two challenge-response messages sent through wireless and ultrasonic channels. The relative delay in the two channels is compared with the ideal theoretic delay, the latter of which  is derived according to a prover's claimed location. Applying the Echo protocol in VANETs requires that the vehicles are equipped with both wireless and ultrasonic hardware for communications.
A location verification protocol with hidden or mobile base stations was presented in \cite{capkun2008secure}. The hidden or mobile base stations can securely estimate the distances to the prover since the locations of the hidden or mobile base stations are assumed unknown to the prover. The distance error between the estimated and claimed locations of a prover is compared with some threshold as a means to verification. This algorithm can be used in VANETs if the locations of some verifiers are not publicly known. In \cite{zhang2008evaluation}, several location verification algorithms were proposed based on power-modulated challenge-response method to detect malicious vehicles. Adopting  power-modulated location verification algorithms to VANETs, would be straightforward if the verifiers in VANETs (vehicle or base stations) possess the ability to adjust their transmit power.

The authors of \cite{ekici2008secure} proposed a probabilistic location verification algorithm for a wireless sensor network (WSN) with high node density. In such networks, the number of hops a packet (sent by a prover) traverses in order to reach a verifier, is shown to be probabilistically dependent on the Euclidean distance between the prover and the verifier. The proposed algorithm in  \cite{ekici2008secure}  verifies a prover's claimed location by checking the correlation between the number of hops and the Euclidean distance (which is calculated based on the prover's claimed location). Assuming a high node density  for the WSN, two location verification algorithms were proposed in \cite{wei2012lightweight}. These algorithms explored the inconsistencies between a prover's claimed location and the verifiers one-hop neighbor's determination that it can hear the prover. Since the algorithms in \cite{ekici2008secure} and \cite{wei2012lightweight}  assumed a high node density for the WSN, applying these algorithms in VANETs requires that the vehicle density is high (suitable for urban scenarios).

Again, we point out that all the works discussed in this last section are just representative (not exhaustive) of location verification algorithms proposed for other wireless networks. We have tried to classify them into three classes, RSS-based, challenge-response based, and high-node-density based . Other classes exist, which again with some thought could be altered in some fashion so as to be applicable to the VANET scenario.

\section{Conclusions} \label{Conclusions}

  In this paper we have outlined generic frameworks for location verification in the context of the VANET scenario. We have also reviewed how much of the exiting literature on location verification with VANETs falls within such frameworks. With the Intelligent Transport System scenario now becoming a key focus of government transportation departments around the world, the deployment of actual VANETs is close to reality. A mission-critical component of such networks will be location verification. As such, the  research reviewed in this paper is likely to be of increasing importance.

\section{Acknowledgment} \label{acknowledgment}

This work has been supported by the University of New
South Wales, and the Australian Research Council, grant DP120102607.

\end{document}